\documentclass[aps,twocolumn]{revtex4}
\pdfoutput=1

\usepackage{graphicx}
\usepackage{times}
\usepackage{amsmath}
\usepackage{amssymb}

\begin{document}
\title{Conformational selection and induced changes along the catalytic cycle of {\em E.~coli} DHFR}

\author{Thomas R.\ Weikl$^{1}$ and David D. Boehr$^2$ \\[0.2cm]
\small $^1$Max Planck Institute of Colloids and Interfaces, Department of Theory 
and Bio-Systems, 14424 Potsdam, Germany \\[0cm]
\small $^2$Department of Chemistry, The Pennsylvania State University,  University Park, Pennsylvania, 16802, USA\\[0cm]
}

\begin{abstract}
Protein function often involves changes between different conformations. Central questions are how these conformational changes are coupled to the binding or catalytic processes during which they occur, and how they affect the catalytic rates of enzymes. An important model system is the  enzyme dihydrofolate reductase (DHFR) from {\em E.~coli}, which exhibits characteristic conformational changes of the active-site loop during the catalytic step and during unbinding of the product. 
In this article, we present a general kinetic framework that can be used (1) to identify the ordering of events in the coupling of conformational changes, binding and catalysis and (2) to determine the rates of the substeps of coupled processes from a combined analysis of NMR $R_2$ relaxation dispersion experiments and traditional enzyme kinetics measurements. We apply this framework to {\em E.~coli} DHFR and find that the conformational change during product unbinding follows a conformational-selection mechanism, i.e.~the conformational change occurs predominantly {\em prior to} unbinding. The conformational change during the catalytic step, in contrast, is an induced change, i.e.~the change occurs {\em after} the chemical reaction. We propose that the reason for these conformational changes, which are absent in human and other vertebrate DHFRs, is robustness of the catalytic rate against large pH variations and changes to substrate/product concentrations in {\em E.~coli}.
\end{abstract}

\maketitle


\section*{INTRODUCTION}

The structural changes in an enzyme are intimately linked to its function. Ligand-exchange events of enzymes through either binding/unbinding or chemical transformation are often accompanied by conformational changes to the protein \cite{Gerstein98,Goh04}. Indeed, the rates of catalysis are frequently limited by the slow conformational changes in enzymes (e.g. \cite{Eisenmesser02,Boehr06a,Watt07}). Optimal enzyme engineering would thus require a deep understanding of both the chemical steps of catalysis and how the structural changes in the enzyme help to facilitate these events \cite{Mandell09,Doucet11,Nagel09}. Of course, one powerful source of functional information comes from the study of enzyme kinetics. However, classical enzyme kinetics have traditionally followed the chemical transformation of substrate to product, and only give indirect evidence about structural changes to the enzyme itself. For example, substrate binding involves the formation of noncovalent interactions between enzyme and substrate, but a conformational change in the enzyme (and/or substrate) may be required to maximize these interactions; traditional enzyme kinetics may not be able to tease out these two separate, but related events. More recent biophysical techniques, including those from single-molecule spectroscopy \cite{Min05,Michalet06,Smiley06} and solution-state nuclear magnetic resonance (NMR) \cite{Palmer04,Mittermaier06,Boehr06b,Henzler07,Loria08}, can monitor structural changes in an enzyme itself more directly.  A synthesis of kinetic information from these recent biophysical experiments and from more traditional ensemble enzyme kinetics would bring a richer understanding of the physical and chemical barriers to catalysis.

\begin{figure}[b]
\begin{center}
\resizebox{0.95 \columnwidth}{!}{\includegraphics{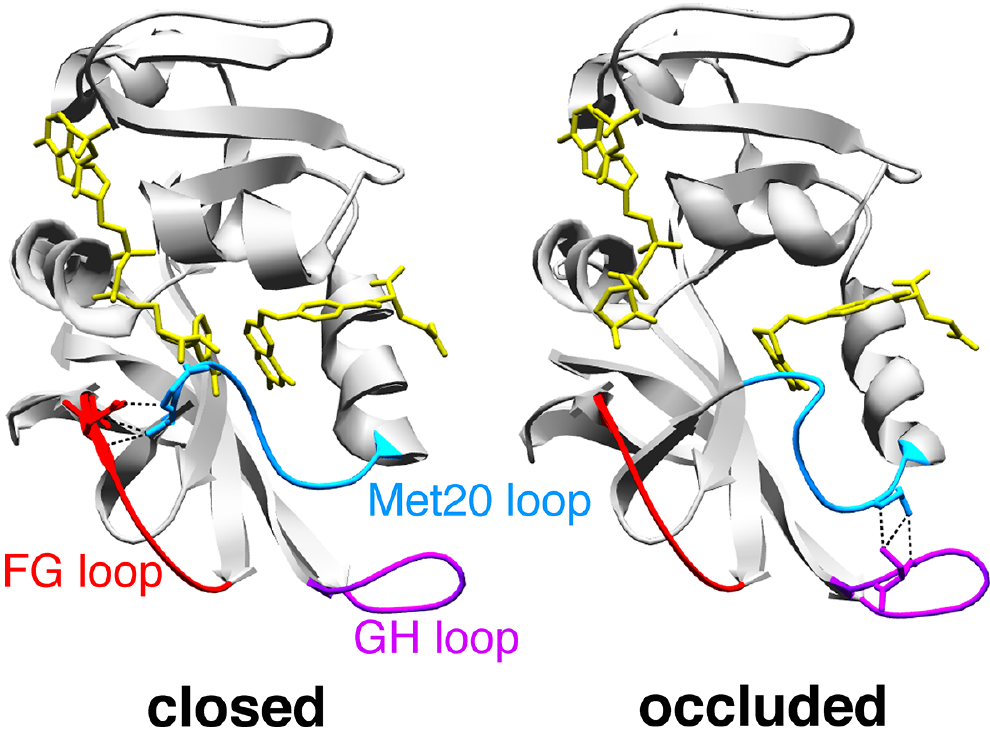}}
\caption{{\em E.~coli} DHFR cycles between closed (left; PDB 1RX2) and occluded (right; PDB 1RX4) conformations. These conformations are dependent on interactions between three loops. In the closed conformation, the active-site Met20 loop (in blue) hydrogen bonds to the FG loop (in red). Attainment of the occluded conformation requires breaking of these hydrogen bonds and formation of new hydrogen bonds between the Met20 and GH (in purple) loops.}  
\label{figure_structures}
\end{center}
\end{figure}

One technique that has brought deep insights into the ``conformational gymnastics" an enzyme undergoes during its catalytic cycle is NMR $R_2$ relaxation dispersion spectroscopy \cite{Palmer04,Mittermaier06,Loria08,Mittermaier09}. This technique measures conformational exchange between two (or more) protein conformations, and can give insight into the kinetics of exchange, the populations of each of the exchanging conformations (i.e.~the thermodynamics), and can give structural information about higher energy conformations, even when the conformational equilibrium is highly skewed (i.e.~less than 5\% of the conformational ensemble is in the higher energy conformation) \cite{Boehr06b,Loria08,Mittermaier09}.  Such methodology has been used to monitor conformational exchange processes on the $\mu$s-ms timescale for all of the intermediates in the catalytic cycle of {\em Echerichia coli} dihydrofolate reductase (DHFR) \cite{Boehr06a,McElheny05}. DHFR catalyzes the reduction of dihydrofolate (DHF) to tetrahydrofolate (THF) using the cofactor NADPH, and has served as an important model system for investigating the interplay of conformational dynamics, binding and catalysis  \cite{Benkovic03,Hammes06,Schnell04,Agarwal02,Thorpe05,Liu07,Chen07,Thielges08}. Intriguingly, the higher energy or ``excited-state" conformations for all the intermediates in DHFR catalysis are structurally similar to either the next and/or previous ``ground-state" conformation in the catalytic cycle of DHFR \cite{Boehr06a}. In other words, the excited-state conformations predict the next structural state of the enzyme. With such rich functional information, {\em E.coli} DHFR is ideal in the development of a kinetic framework that encompasses both the traditional pre-steady state enzyme kinetics and the kinetics on the structural changes in the enzyme gleaned from the NMR $R_2$ relaxation dispersion experiments.
	
The $R_2$ relaxation dispersion results from DHFR also prompted the suggestion that ligand-exchange events on this enzyme, and likely other enzymes \cite{Loria08}, can operate through a ``conformational selection" mechanism. In conformational selection \cite{Tsai01}, conformational changes occur {\em predominantly prior} to a binding process, an unbinding process, or a catalytic process; the ligands appear to select a conformation from the protein ensemble for binding, unbinding, or catalysis \cite{Bosshard01,James03,Beach05,Tobi05,Boehr06a,Kim07,Qiu07,Lange08,Braz10,Masterson10}. In contrast, ``induced change" \cite{Koshland58} proposes that the conformational change occurs {\em predominantly after} a binding, unbinding, or catalytic process \cite{Sullivan08,Elinder10,Fieulaine11,Copeland11}. Critically, the timing of the ligand interaction and the structural changes to the enzyme is not trivial; the processes of conformational selection and induced change have important biological consequences in terms of substrate promiscuity, enzyme regulation, protein evolution, and biological information storage \cite{Boehr06a,Boehr11,Boehr09}. 

In this article, we present a general kinetic framework that can be used (1) to distinguish conformational-selection from induced-change processes, and (2) to determine the rates of the physical substeps (conformational changes) and chemical substeps (binding/unbinding, or chemical reactions) of these processes from a combined analysis of NMR $R_2$ relaxation dispersion  and traditional enzyme kinetics experiments. We apply this framework here to {\em E.~coli} DHFR and derive extended catalytic cycles of this enzyme that specify the ordering of events in the coupling of conformational changes, binding and catalysis. We focus on the major conformational changes of {\em E.~coli} DHFR  that occur during the catalytic step (from ``closed" to ``occluded", see fig.~\ref{figure_structures}) and during unbinding of the product THF (from ``occluded" back to ``closed"). We find that product unbinding occurs {\em via} conformational selection, whereas the conformational change during the catalytic step is an induced-change process (see Results section). Our extended cycles help to explain the role of the conformational changes for the function of {\em E.~coli} DHFR (see Discussions section). 

The methods employed here to analyze the catalytic step and product unbinding of  {\em E.~coli} DHFR can be generalized to binding, unbinding, and catalytic processes of other protein systems. Our methods are based on general expressions for the effective rates of processes that involve two or more substeps, and on a mutational analysis of these effective rates. In this analysis, we focus on mutations distal to the binding site that mainly affect the conformational equilibrium of the protein. We find that conformational-selection processes are sensitive to such distal mutations  because these processes involve a change to a low-populated, excited-state conformation,  and because the equilibrium probability and excitation rate of this conformation depend on the conformational free-energy differences. In contrast, induced-change processes involve a conformational relaxation into a new ground-state after a binding, unbinding or catalytic event, which is rather insensitive to changes in conformational free-energy differences, provided (i) the conformational relaxation is fast compared to the rates of the chemical substep \cite{Weikl09}, or (ii) the transition-state for the conformational exchange is close in free energy and structure to the excited protein conformation. The analysis of the effect of distal mutations thus can provide the basis for a simple diagnostic to identify conformational-selection {\em versus} induced-change processes.

\begin{figure}[t]
\begin{center}
\resizebox{\columnwidth}{!}{\includegraphics{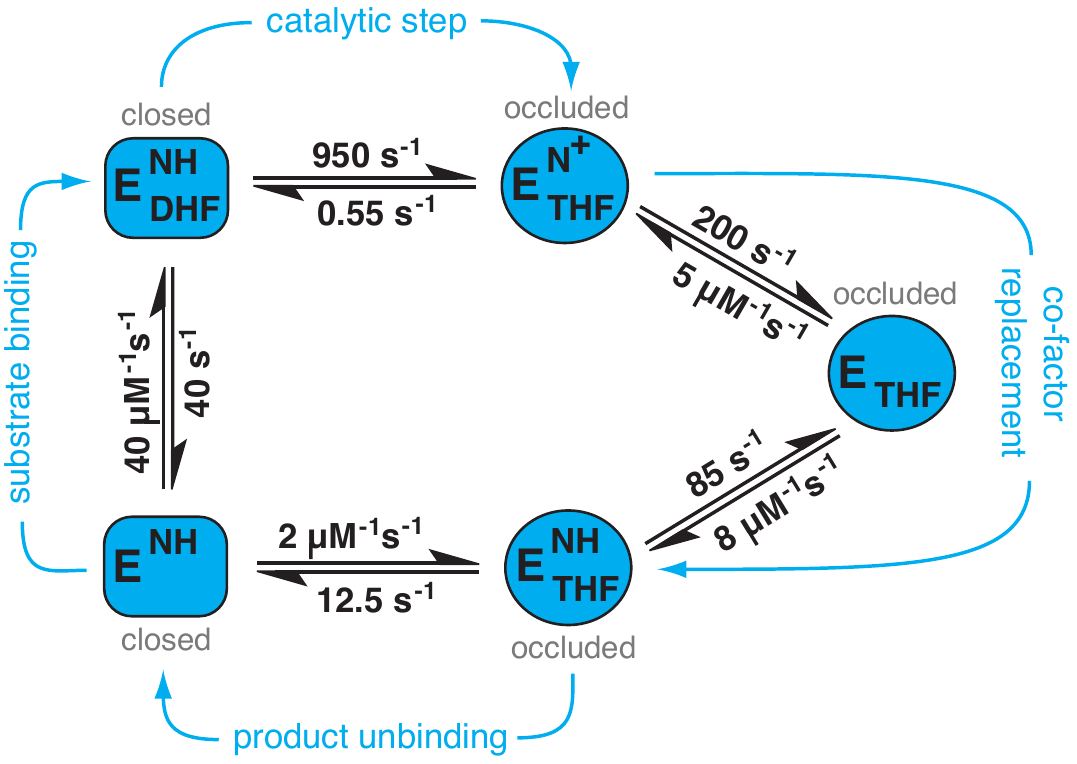}}
\caption{Catalytic cycle of the enzyme dihydrofolate reductase (DHFR) from {\em E.~coli} \cite{Fierke87}. DHFR catalyzes the reduction of dihydrofolate (DHF) to tetrahydrofolate (THF), using NADPH (NH) as a cofactor. The enzyme (E) cycles through five major intermediates\cite{Fierke87,Benkovic03,Hammes06}.  X-ray crystal structures of the enzyme in these five intermediates indicate characteristic conformational changes \cite{Sawaya97}. In the intermediate states $\text{E}^\text{NH}$ and $\text{E}^\text{NH}_\text{DHF}$, the active-site loop is predominantly `closed' over the reactants (see fig~\ref{figure_structures}). In the other three intermediates, this loop `occludes' (i.e.~protrudes  into) the active site. The conformational changes occur during the catalytic step, and during unbinding of the product THF. In this so-called pH-independent scheme \cite{Fierke87}, the forward rate of the catalytic step is the maximal rate at low pH, and the reverse rate of this step is the maximal rate at high pH (see text).}
\label{figure_WT_cycle}
\end{center}
\end{figure}
\begin{figure*}
\begin{center}
\resizebox{1.4\columnwidth}{!}{\includegraphics{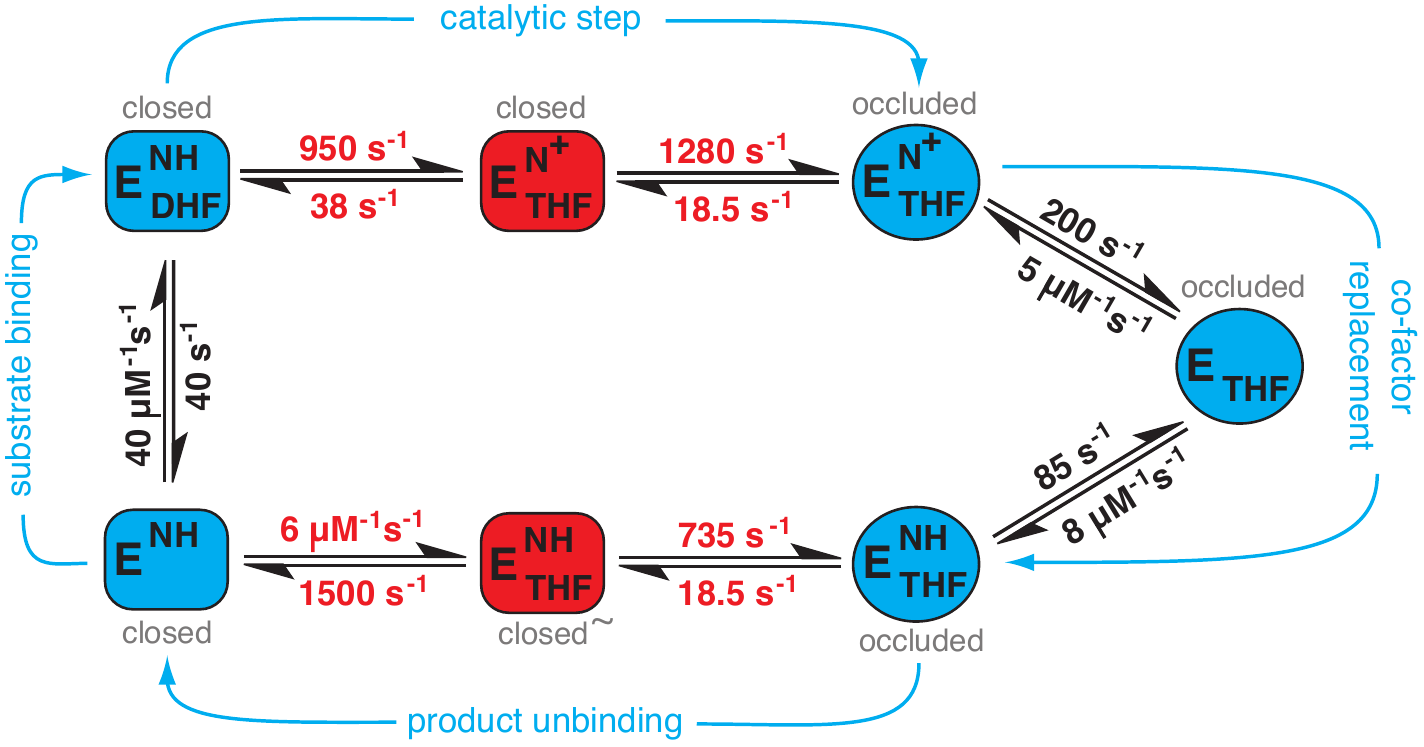}}
\caption{Extended catalytic cycle of the enzyme DHFR from {\em E.~coli}. Excited-state conformations are shown in red, ground-state conformations in blue. On this extended cycle, the catalytic step is decomposed into the actual chemical substep in the closed conformation of the enzyme required for hydride transfer, and a physical substep in which the enzyme conformation changes from closed to occluded. The given rates for the chemical substep follow from eqs.~(\ref{kmin}) and (\ref{kf}) and the experimentally determined maximal rates for the overall catalytic step shown in fig.~\ref{figure_WT_cycle}. The transition rates between the ground-state and excited-state conformations have been determined in NMR relaxation experiments \cite{Boehr06a}. Our analysis of the product-unbinding kinetics and NMR relaxation experiments \cite{Boehr06a} indicate that the conformational change from the occluded to the closed conformation occurs largely prior to the unbinding of the product THF. Along our extended cycle, the product unbinds from an excited state with a conformation similar to the closed conformation (denoted as `closed$^{\sim}$'). The given binding and unbinding rates of THF follow from eqs.~(\ref{cmin}) and (\ref{cplus}) and the overall rates for the product binding and unbinding process shown in fig.~\ref{figure_WT_cycle}. 
}
\label{figure_WT_cycle_extended}
\end{center}
\end{figure*}
%

\section*{RESULTS}

%
\subsection*{Structural changes in the catalytic cycle of {\em E.~coli} DHFR}

The catalytic cycle of {\em E.~coli} DHFR involves five major intermediates (see fig.~\ref{figure_WT_cycle}). X-ray crystal structures indicate an `occluded' ground-state conformation of the active-site Met20 loop (residues 9-24) in the three intermediates with bound product THF, and a `closed' conformation of this loop in the other two intermediates  \cite{Sawaya97}. When the Met20 loop is in an occluded conformation, its central region (residues Met16 and Glu17) is flipped into the cofactor binding pocket such that the nicotinamide ring of the cofactor is sterically occluded from binding. However upon binding cofactor, residues Met16 and Glu17 are turned away from the nicotinamide binding pocket, residues Met16-Ala19 zip up into a type III' $\beta$-hairpin, and the side-chains of Asn18 and Met20 close down over the pterin ring of the substrate DHF to form the closed conformation.  Only the closed conformation allows proximal positioning of the cofactor and substrate reactive centers and therefore it is the closed conformation that must be adopted in the Michaelis complex. The closed and occluded Met20 loop conformations are stabilized by unique hydrogen bonding networks to the FG (116-132) and GH (142-150) loops, respectively (see fig.~\ref{figure_structures}). Changes to the amino acids in the FG or GH loops that make contact with the Met20 loop can severely alter the equilibrium between the closed and occluded conformations of the Met20 loop. The mutant G121V, for example, has the occluded conformation as ground state in all five intermediates along the catalytic cycle \cite{Venkitakrishnan04}, while the mutant S148A seems to have the closed conformation of the active-site loop as ground state in all intermediates \cite{Bhabha11}. These amino acid substitutions have functional consequences for both the chemical step and the THF product release step that are accompanied by closed-occluded conformational changes.

\subsection*{The catalytic step of {\em E.~coli} DHFR operates through an induced-change mechanism}

Two steps in the catalytic cycle of {\em E.~coli} DHFR are accompanied by major structural changes in the enzyme -- the catalytic step itself and the release of the product THF from the ternary complex of the enzyme with THF and NADPH. Division of the catalytic cycle into chemical (including substrate binding/product release and chemical transformation of substrate to product) and physical substeps requires an investigation of both steps. Hydride transfer in the chemical step is only possible in the closed conformation, thus any conformational change in the enzyme through the catalytic step must occur after THF production i.e.~through an induced-change process. In the following, we apply our kinetic framework to determine the rates of the chemical and physical substeps of this induced-change process from experimental data.

The catalytic step of wild-type DHFR is associated with a conformational change from the closed conformation (cE) to the occluded conformation (oE) of the enzyme (see fig.~\ref{figure_WT_cycle}):
\begin{equation}
\text{cE}_\text{DHF}^\text{NH} \;
\begin{matrix} 
\text{\footnotesize $k_f$}\\[-0.05cm]
\text{\Large $\rightleftharpoons$}\\[-0.15cm]
\text{\footnotesize $k_r$} \\[0.05cm]
\end{matrix}
 \; \text{oE}_\text{THF}^\text{N+} 
 \label{catalytic_process}
\end{equation}
The forward and reverse rates $k_f$ and $k_r$ for the catalytic step are pH-dependent since the catalyzed reaction $\text{DHF}+\text{NH}+\text{H}^{+}\rightleftharpoons \text{THF} + \text{N}^{+}$ requires an additional proton H$^{+}$ in the forward direction. The forward rates $k_f$ from enzyme kinetics measurements at different pH  are well described by \cite{Fierke87}
\begin{equation}
k_f(\text{pH}) \simeq k_\text{max}/\left(1 + 10^{\text{pH}-\text{pKa}}\right)
\label{kf(pH)}
\end{equation}
with the maximal forward rate $k_\text{max} = 950 \pm 50$ s$^{-1}$ at low pH and $\text{pKa}=6.5 \pm 0.1$ at 25$^\circ$C. The reverse rate $k_r$ attains its maximal value $0.55 \pm 0.1$ s$^{-1}$ at high pH \cite{Fierke87} and has the value $k_r\simeq 0.03$ s$^{-1}$ at pH 7 \cite{Cameron97}. 

We now break down the catalytic step into its chemical and physical substeps. Since the hydride transfer from NH to DHF is only possible in the closed conformation (see above), the chemical hydride-transfer substep occurs before the physical substep, the conformational change from the closed to the occluded conformation:
\begin{equation}
\text{cE}_\text{DHF}^\text{NH} \;
\begin{matrix} 
\text{\footnotesize $k_{+}$}\\[-0.05cm]
\text{\Large $\rightleftharpoons$}\\[-0.15cm]
\text{\footnotesize $k_{-}$} \\[0.05cm]
\end{matrix}
 \; \text{cE}_\text{THF}^\text{N+} \;
 \begin{matrix} 
\text{\footnotesize $k_{co}$}\\[-0.05cm]
\text{\Large $\rightleftharpoons$}\\[-0.15cm]
\text{\footnotesize $k_{oc}$} \\[0.05cm]
\end{matrix}
\; \text{oE}_\text{THF}^\text{N+}
\label{catalysis_DHFR}
\end{equation}
The occluded conformation $\text{oE}_\text{THF}^\text{N+}$ is the ground state of the enzyme with bound products THF and N$^+$, and the closed conformation $\text{cE}_\text{THF}^\text{N+}$ is an excited state. The rates $k_{oc}$ and $k_{co}$ for this conformational transition have been measured in NMR $R_2$ relaxation dispersion experiments. At 27$^\circ$C, these rates are $k_{oc} = 18.5 \pm 1.5$ s$^{-1}$  and $k_{co} = 1280 \pm 50$ s$^{-1}$  \cite{Boehr06a}. The $R_2$ relaxation dispersion experiments indicate that the conformational dynamics of the active-site loop does not depend on the pH \cite{Boehr08}.

The rates $k_{+}$ and $k_{-}$ for the chemical substep in the closed conformation can be calculated from (i) the rates $k_r$ and $k_f$ for the overall catalytic step determined in enzyme kinetics experiments, and (ii) the rates $k_{oc}$ and $k_{co}$ for the physical substep determined in $R_2$ relaxation dispersion experiments. Our calculations are based on general results for the effective rates of processes with two (or more) substeps (see Appendix). We first focus on the reverse direction. The effective rate $k_r$ for the reverse hydride transfer depends on $k_{oc}$, $k_{co}$ and $k_{-}$. Since the excitation rate $k_{oc}$ is much smaller than the ground-state relaxation rate $k_{co}$, the effective reverse rate is $k_r \simeq k_{oc} k_{-}/(k_{co} + k_{-})$ (see eq.~(\ref{kAC}) in Appendix), which leads to
\begin{equation}
k_{-} \simeq \frac{k_{co} k_r}{k_{oc} - k_r} \simeq \frac{k_{co}}{k_{oc}} k_r \simeq (69\pm6)\cdot k_r 
\label{kmin}
\end{equation}
since $k_{oc}$ is much larger than $k_r$. We have inserted here the values for the conformational transition rates $k_{oc}$ and $k_{co}$ measured in the NMR $R_2$ relaxation dispersion experiments. For the experimentally determined maximal effective rate $k_r = 0.55 \pm 0.1$ s$^{-1}$ at high pH, we obtain $k_{-} = 38\pm 8$ s$^{-1}$ as maximal reverse rate. 

The effective rate $k_f$ in the forward direction depends on the rates $k_{+}$, $k_{-}$, and the relaxation rate $k_{co}$ into the ground state with bound products THF and N$^+$. Because $k_{co}$ is much larger than $k_{-}$, the effective rate in the forward direction is either $k_{+}$ or $k_{co}$, depending on which of these two rates is smaller and, thus, rate-limiting (see eq.~(\ref{kAC})). Since the ground-state relaxation rate $k_{co}$ is larger than the maximum effective rate in the forward direction, we have
\begin{equation}
k_f \simeq k_+
\label{kf}
\end{equation}
This equation is consistent with eq.~(\ref{kmin}) because the equilibrium condition $k_f/k_r \simeq (k_{oc}/k_{co}) (k_{+}/k_{-})$ is fulfilled. Our decomposition of the catalytic step into its physical and chemical substeps is included in the extended catalytic cycle of {\em E.~coli} shown in fig.~\ref{figure_WT_cycle_extended}.

\subsection*{Product release from {\em E.~coli} DHFR may occur through three potential routes}	

\label{section_product_unbinding}

Product release is the rate-limiting step in the steady-state catalytic cycle of DHFR  for pH values up to the preferred internal pH of {\em E.~coli} (see Discussion section below), and so, has a special importance in DHFR function. In contrast to the catalytic step, the unbinding of the product may occur along three possible routes on which the conformational change from the occluded to closed conformations occurs either (1) prior to, (2) after, or (3) partly prior to and partly after the unbinding event (see Figures 4 and 5). By calculating the effective rates along these three routes and by determining the effect of distal mutations on the off-rates, we can identify which of these routes is the product unbinding route of {\em E.~coli} DHFR. It should be kept in mind that these three routes can be generalized to other protein systems and are applicable to any ligand (small molecule and/or large macromolecule) binding/unbinding event, { including binding events that involve folding of a ligand \cite{Kiefhaber12,Wright09}. An implicit assumption of these routes is that the chemical and physical substeps can be separated in time.} 

{\bf Route 1:} We first consider the conformational-selection unbinding route
\begin{equation}
\text{oE}_\text{THF}^\text{NH} \;
\begin{matrix} 
\text{\footnotesize $b_{oc}$}\\[-0.05cm]
\text{\Large $\rightleftharpoons$}\\[-0.15cm]
\text{\footnotesize $b_{co}$} \\[0.05cm]
\end{matrix}
 \; \text{cE}_\text{THF}^\text{NH} 
\begin{matrix} 
\text{\footnotesize $c_{-}$}\\[-0.05cm]
\text{\Large $\rightleftharpoons$}\\[-0.15cm]
\text{\footnotesize $c_{+}\mbox{[THF]}$} \\[0.05cm]
\end{matrix}
\; \text{cE}^\text{NH}
\label{route1}
\end{equation}
along which the conformational change precedes product unbinding. We assume here and below that the concentration of the product THF is significantly larger than the concentration of the enzyme, which implies pseudo-first-order kinetics with approximately constant product concentration and, thus, constant binding rate $c_{+}[\text{THF}]$. 

\begin{figure}[b]
\begin{center}
\resizebox{0.65\columnwidth}{!}{\includegraphics{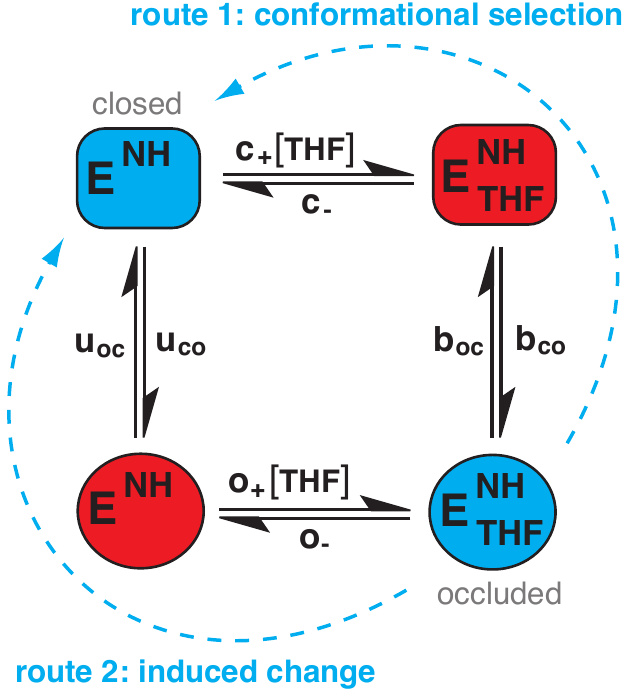}}
\caption{Possible unbinding routes of the product THF from wildtype {\em E.~coli} DHFR. On the conformational-selection route 1, the conformational change from the occluded to the closed conformation of the active-site loop occurs {\em prior to} product unbinding. On the induced-change route 2, the conformational change occurs {\em after} unbinding. 
}
\label{figure_routes1and2}
\end{center}
\end{figure}
\begin{figure*}
\begin{center}
\resizebox{1.2\columnwidth}{!}{\includegraphics{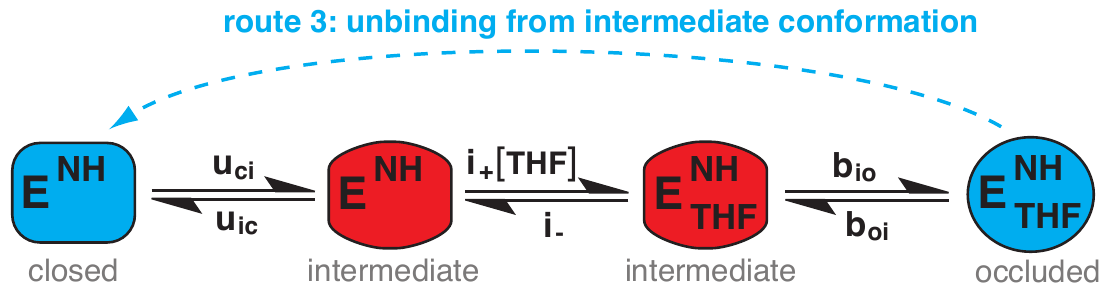}}
\caption{Possible unbinding route of the product THF from wildtype {\em E.~coli} DHFR. Along this route, the product unbinds from an intermediate conformation. 
}
\label{figure_route3}
\end{center}
\end{figure*}

The effective off-rate $k_\text{off}^{(1)}$ along route 1 depends on the rates $b_{oc}$ and $b_{co}$ of the conformational transitions and on the unbinding rate $c_{-}$ from the closed conformation. Since the excitation rate $b_{oc}$ is much smaller than the relaxation rate $b_{co}$ into the product-bound ground state, the effective off-rate is (see eq.~(\ref{kAC})) 
\begin{equation}
k_\text{off}^{(1)} \simeq \frac{b_{oc}\, c_{-}}{b_{co} + c_{-}}
\label{koff1}
\end{equation}
The effective on-rate constant $k_\text{on}^{(1)}$ on route 1 depends on the binding rate constant $c_{+}$ and the unbinding rate $c_{-}$ in the closed conformation, and on the relaxation rate $b_{co}$ into the product-bound ground state.  If the relaxation rate $b_{co}$ is much larger than the binding rate $c_{+}[\text{THF}]$, the effective on-rate constant is (see eq.~(\ref{kAC})) 
\begin{equation}
k_\text{on}^{(1)} \simeq \frac{b_{co}\,c_{+} }{b_{co} + c_{-}}
\label{kon1}
\end{equation}
The eqs.~(\ref{koff1}) and (\ref{kon1}) fulfill the equilibrium condition $k_\text{on}^{(1)}/k_\text{off}^{(1)} = (b_{co}/b_{oc})(c_{+}/c_{-})$.

Amino acid substitutions in the FG and GH loops of {\em E.~coli} DHFR, such as G121V and S148A, affect the conformational equilibrium between closed and occluded states, but these residues do not make direct interactions with the THF product or NADPH cofactor themselves. These distal mutations thus should mainly affect the free-energy difference  $\Delta G_{co}$ between the closed and occluded conformation. We will show in the following that the analysis of distal mutations in the context of the three routes of product unbinding can be used to separate out the chemical and physical substeps of product release and give insight into the ordering of the events. 

\begin{figure}[t]
\begin{center}
\resizebox{0.65\columnwidth}{!}{\includegraphics{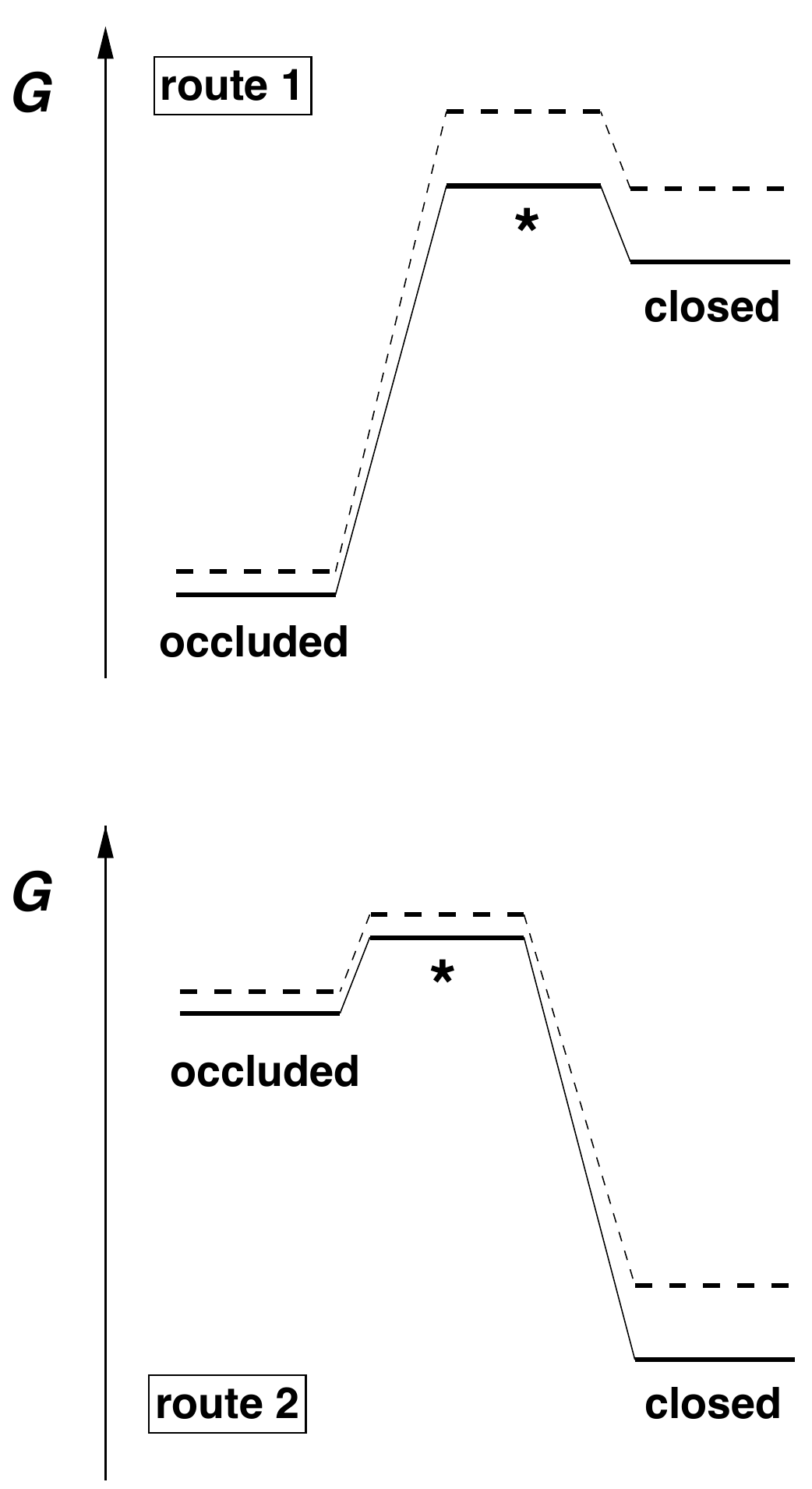}}
\caption{Energy landscapes for the conformational changes along route 1 and route 2 of fig.~\ref{figure_routes1and2} in transition-state theory. The full lines indicate free-energy levels for the wildtype, the dashed lines the free-energy levels for a mutant. Along route 1, the conformational change occurs in the product-bound state in which the occluded conformation is lower in free energy than the closed conformation. In the landscape for route 1, the free-energy difference between the transition state and the closed conformation is small compared to the difference between the transition state and the occluded conformation. According to the Hammond-Leffler postulate, the structure of the transition state then resembles the structure of the closed state, which implies that mutations have a similar effect on these two states. Along route 2, the conformational change occurs in the unbound state in which the closed conformation is lower in free energy. Since the transition state now is much closer in free energy to the occluded state, mutations have a similar effect on these two states  according to the Hammond-Leffler postulate.
  }
\label{figure_ts}
\end{center}
\end{figure}

In transition-state theory, the conformational transition rates $b_{oc}$ and $b_{co}$ depend on free-energy differences to the transition state:
\begin{equation}
b_{oc} = b_o e^{-\Delta G_{*o}/RT} \text{~~and~~} b_{co} = b_o e^{-\Delta G_{*c}/RT}
\label{tst}
\end{equation}
Here, $\Delta G_{*o}$ is the free energy difference between the transition state and the occluded conformation, and $\Delta G_{*c}$  the free-energy difference between the transition state and the closed conformation. Because of $b_{oc}\ll b_{co}$, it seems not unreasonable to assume that the free-energy difference $\Delta G_{*c}$ is significantly smaller than $\Delta G_{*o}$, i.e.~that the transition state is significantly closer in free energy to the closed conformation (see fig.~\ref{figure_ts}). According to the Hammond-Leffler postulate, the structure of the transition state then resembles the structure of the closed conformation\cite{Hammond55,Leffler53}. Distal mutations thus should have a similar effect on the transition state and the closed conformation, which implies $\Delta\Delta G_{*o} \simeq \Delta\Delta G_{co}$ and  $\Delta\Delta G_{*c} \ll \Delta\Delta G_{co}$ for the mutation-induced changes of the free-energy differences. This leads to 
\begin{equation}
b_{oc}^\prime/ b_{oc} \simeq e^{-\Delta\Delta G_{co}/RT} \text{~~and~~} b_{co}^\prime \simeq b_{co}
\label{ts}
\end{equation}
where the prime indicating rates of the mutant,  and $\Delta\Delta G_{co} = \Delta G^\prime_{co} - \Delta G_{co}$ is the mutation-induced change of free-energy difference between the conformations.  We have assumed here that the pre-exponential factor $b_o$ in eq.~(\ref{tst}) its not affected by the mutation. According to eq.~(\ref{ts}), mutational shifts of the conformational free-energy difference affect the excitation rate $b_{oc}$, but not the relaxation rate $b_{co}$.

If we neglect the effect of the distal mutations on the unbinding rate $c_{-}$ from the closed conformation, we obtain
\begin{equation}
\frac{k_\text{off}^{\prime\, (1)}}{k_\text{off}^{(1)}} \simeq \frac{b_{oc}^\prime}{b_{oc}} \simeq  e^{-\Delta\Delta G_{co}/RT}
\label{koff1ratio}
\end{equation}
from eqs.~(\ref{koff1}) and (\ref{ts}). In the special case  $c_{-} \ll b_{co}$ of a conformational relaxation rate $b_{co}$ that is much larger than the unbinding rate $c_{-}$, eq.~(\ref{koff1ratio}) also follows directly from eq.~(\ref{koff1}) and $b_{oc}/b_{co} = e^{-\Delta G_{co}/RT}$ for $c_{-}^\prime \simeq c_{-}$, because we have $k_\text{off}^{(1)} \simeq (b_{oc}/b_{co}) c_{-}$ in this case \cite{Weikl09}.  However, this special case does not seem to apply to DHFR product unbinding (see below). Eq.~(\ref{koff1ratio}) simply states  that the ratio of the mutant and wildtype off-rates for the conformational-selection process depends on the shift $\Delta\Delta G_{co}$ of the free energy-difference between the closed and occluded conformations induced by a distal mutation.

{\bf Route 2:}  On the induced-change unbinding route, the conformational change occurs after the unbinding of the product THF: 
\begin{equation}
\text{oE}_\text{THF}^\text{NH}
\begin{matrix} 
\text{\footnotesize $o_{-}$}\\[-0.05cm]
\text{\Large $\rightleftharpoons$}\\[-0.15cm]
\text{\footnotesize $o_{+} [\text{THF}]$} \\[0.05cm]
\end{matrix}
 \;
 \text{oE}^\text{NH}
 \;
 \begin{matrix} 
\text{\footnotesize $u_{oc}$}\\[-0.05cm]
\text{\Large $\rightleftharpoons$}\\[-0.15cm]
\text{\footnotesize $u_{co}$} \\[0.05cm]
\end{matrix}
\;\text{cE}^\text{NH} 
\label{route2}
\end{equation}
Along this unbinding route, the product THF unbinds from the occluded ground-state conformation, which induces the conformational change into the closed conformation. 

The effective off-rate rate $k_\text{off}^{(2)}$ for the induced-change route 2 depends on the unbinding rate $o_{-}$ from the occluded conformation, the rebinding rate $o_{+} [\text{THF}]$, and the relaxation rate $u_{oc}$ into the unbound ground state. If the relaxation rate $u_{oc}$ is much larger than the rebinding rate  $o_{+} [\text{THF}]$, the effective unbinding rate $k_\text{off}^{(2)}$ is either $o_{-}$ or $u_{oc}$, depending on which of the two rates is smaller (see eq.~(\ref{kAC})). { Because of  $u_{co}\ll u_{oc}$, we assume again that the transition state for the conformational exchange is significantly closer in free energy and structure to the excited-state conformation (see fig.~\ref{figure_ts}), which implies that distal mutations should mainly affect the excitation rate $u_{co}$ 
rather than the  relaxation rate $u_{oc}$, as in eq.~(\ref{ts}).  Since the effective off-rate $k_\text{off}^{(2)}$ does not depend on the excitation rate $u_{co}$, we obtain
\begin{equation}
\frac{k_\text{off}^{\prime\, (2)}}{k_\text{off}^{(2)}}  \simeq 1 
\label{koff2ratio}
\end{equation}
if we neglect the effect of the distal mutation on the binding rate constant $o_{+}$ and unbinding rate $o_{-}$ in the occluded conformation}.

In contrast to the conformational-selection process, the ligand off-rate for the induced change-process is not affected by a distal mutation, according to eq.~(\ref{koff2ratio}). The eqs.~(\ref{koff1ratio}) and (\ref{koff2ratio}) thus provide the basis for a simple diagnostic to identify conformational-selection {\em versus} induced-change processes. In the next section, we will apply this diagnostic to experimental data for the product unbinding rates of wildtype and mutants of {\em E.~coli} DHFR from the ternary complex $\text{E}_\text{THF}^\text{NH}$, in combination with calculations of mutation-induced stability changes (see fig.~\ref{figure_data}). We focus here on off-rates because corresponding data for on-rates are not available for DHFR. However, a corresponding diagnostic also holds for mutation-induced changes of on-rates \cite{Weikl09}. 

%

\begin{table*}
\caption{Calculated mutation-induced stability changes for DHFR conformations}
\label{table_conformations}
\begin{center}
\begin{tabular}{l | c | c | cccccc}
state & PDB file & active-site loop & D122A &  D122N  & D122S & G121V & M42W & S148A  \\
\hline
$\text{E}^\text{NH}$                        & 1RX1 & closed   & 0.22 & -0.08 & -0.52  & 1.13 & -1.58 & 0.34 \\ 
$\text{E}^\text{NH}_\text{DHF}$      & 1RX2  & closed & -0.28 & -0.79 & -0.45  & 2.15 & 0.41 & -0.26 \\
$\text{E}^{\text{N}^+}_\text{THF}$   & 1RX4 & occluded & -0.42  & -0.09 & -0.25 & -0.26 & -0.01 & 1.04 \\
$\text{E}_\text{THF}$                       & 1RX5 & occluded & -1.15  & -0.63 & -1.22  & -0.74 & -0.20 & 0.31\\
$\text{E}^\text{NH}_\text{THF}$       & 1RX6 & occluded & -1.03  & -0.80 & -0.79  & -0.85 & -0.16 & 0.79 \\
\hline
\end{tabular}
\end{center}
\begin{flushleft}
Stability changes $\Delta\Delta G$ for DHFR mutants in units of kcal/mol calculated with the program Concoord/PBSA \cite{Benedix09}. The calculation error can be estimated as the standard deviation of 1.04 kcal/mol between experimentally determined and calculated stability changes for 582 mutations of seven proteins \cite{Benedix09}. The given values are averages over five independent calculations with Concoord/PBSA. The standard deviation for the five calculations is on average 0.38 kcal/mol. The mutation-induced stability change is the difference $\Delta\Delta G = \Delta G^\prime - \Delta G $ between the stability $\Delta G^\prime$ of the mutant and the stability $\Delta G$ of the wildtype. The stability $\Delta G$ is the free energy difference between a folded, native conformation and the denatured state. Atoms of the ligand molecules are discarded from the PDB files prior to the calculations.
\end{flushleft}
\end{table*}

{\bf Route 3:} Our diagnostic to identify conformational-selection and induced-change processes can be extended to route 3 on which the product unbinds from an intermediate conformation $i$ between the occluded and the closed conformation:
\begin{equation}
\text{oE}_\text{THF}^\text{NH} \;
\begin{matrix} 
\text{\footnotesize $b_{oi}$}\\[-0.05cm]
\text{\Large $\rightleftharpoons$}\\[-0.15cm]
\text{\footnotesize $b_{io}$} \\[0.05cm]
\end{matrix}
 \; \text{iE}_\text{THF}^\text{NH}  
\begin{matrix} 
\text{\footnotesize $i_{-}$}\\[-0.05cm]
\text{\Large $\rightleftharpoons$}\\[-0.15cm]
\text{\footnotesize $i_{+} [\text{THF}]$} \\[0.05cm]
\end{matrix}
 \;
 \text{iE}^\text{NH}
 \;
 \begin{matrix} 
\text{\footnotesize $u_{ic}$}\\[-0.05cm]
\text{\Large $\rightleftharpoons$}\\[-0.15cm]
\text{\footnotesize $u_{ci}$} \\[0.05cm]
\end{matrix}
\;\text{cE}^\text{NH} 
\label{route3}
\end{equation}
Along this route, a conformation-selection step from the bound ground state $\text{oE}_\text{THF}^\text{NH}$ to the intermediate state 
 $\text{iE}^\text{NH}$ is followed by an induced-change step into the unbound ground state $\text{cE}^\text{NH}$. The routes 1 and 2 can be seen as limiting cases of route 3 with `intermediates' that are identical with the closed or occluded conformation, respectively (see also \cite{Wlodarski09,Csermely10}).

The effective unbinding rate of route 3 is (see eq.~(\ref{kAD})):
\begin{eqnarray}
k_\text{off}^{(3)}&\simeq& \frac{b_{oi} u_{ic} i_{-}}{b_{io} (u_{ic} + i_{+}[\text{THF}])+ u_{ic} i_{-} } \\
\label{koff3full}
&\simeq& \frac{b_{oi} i_{-}}{b_{io} + i_{-}  } \text{~~for~~} u_{ic} \gg i_{+}[\text{THF}]
\label{koff3}
\end{eqnarray}
For a relaxation rate $u_{ic}$ into the unbound ground state that is much larger than the rebinding rate $i_{+}[\text{THF}]$ of the unbound excited state, the unbinding process thus is {\em dominated by the conformational-selection step} since $k_\text{off}^{(3)}$ is identical with the effective rate from the bound ground state $\text{oE}_\text{THF}^\text{NH}$ to the intermediate state $\text{iE}^\text{NH}$ in this case (see eq.~(\ref{koff1}) for comparison). In analogy to eq.~(\ref{koff1ratio}) for conformation-selection unbinding, the effect of a distal mutation on the unbinding rate therefore can be characterized by
\begin{equation}
\frac{k_\text{off}^{\prime\, (3)}}{k_\text{off}^{(3)}}  \simeq \frac{b_{oi}^\prime}{b_{oi}} \simeq  e^{-\Delta\Delta G_{io}/RT}  \text{~~for~~} u_{ic} \gg i_{+}[\text{THF}]
\label{koff3ratio}
\end{equation}
where $\Delta\Delta G_{io}$ is the mutation-induced change of the free-energy difference between the intermediate and the occluded conformation. On route 3, the ratio of the mutant and wildtype off-rates thus depends on the shift $\Delta\Delta G_{io}$ of the free-energy difference between the intermediate and the occluded conformation. 

\subsection*{Mutational analysis of DHFR indicates that THF release operates through a conformational-selection mechanism}

Based on the above analysis, we can identify whether a ligand-exchange event operates through a conformational-selection or induced-change process by analyzing experimental data for the ligand off-rates (or on-rates) for a series of enzyme mutants, in which the amino acid substitutions primarily alter the free-energy difference between the conformations before and after binding. Benkovic and co-workers have measured the effect of six single-site mutations distal to the binding site on the product unbinding rate of DHFR \cite{Cameron97,Miller98c,Miller01,Rajagopalan02}. Four of the mutations decrease the product unbinding rate, while two of the mutations increase the unbinding rate (see table \ref{table_product_unbinding}). Since most of the mutated residues are in contact with the active-site loop, the effect of the distal mutations on the unbinding kinetics seems to result predominantly from a shift of the conformational equilibrium between the closed and occluded conformation of this loop.

We have calculated the mutation-induced stability changes of the ground-state conformations along the catalytic cycle of DHFR for the six distal mutations (see table \ref{table_conformations}). These ground-state conformations correspond to the crystal structures of the five major intermediates for wildtype {\em E.~coli} DHFR. We have used the program Concoord/PBSA \cite{Benedix09}, which is one of the most reliable programs for the calculation of  mutation-induced stability changes \cite{Potapov09,Kellogg11}, and particularly suited for insertions of larger amino acids as in the mutation G121V since the flexibility of the protein backbone is taken into account. { A central assumption of our calculations is that the distal mutations only affect the conformational free energy of the enzyme but not the binding free energy of the ligands,  which are ``removed" from the crystal structures in the Concoord/PBSA calculations.} For the mutation G121V, we obtain a destabilization of the closed ground-state conformations of the intermediates  $\text{E}^\text{NH}$ and $\text{E}^\text{NH}_\text{DHF}$, and a stabilization of the occluded ground-state conformations of the intermediates  $\text{E}^\text{N+}_\text{THF}$, $\text{E}_\text{THF}$, and $\text{E}^\text{NH}_\text{THF}$ { with respect to the denatured state of the enzyme} (see table \ref{table_conformations}). In agreement with the NMR experiments \cite{Venkitakrishnan04}, our calculations thus indicate a stabilization of occluded conformations relative to closed conformations for G121V. These findings make qualitative sense in light of the protein structure. The larger Val side chain may interfere with the H-bond interactions between Asp122 and Gly15/Glu17 that would act to destabilize the closed conformation. For the mutation S148A, our calculations indicate a destabilization of the occluded ground-state conformations, and a rather neutral effect on the closed ground-state conformations. The loss of the H bond between Ser148 and Asn23 would act to destabilize the occluded state, which is observed in the NMR experiments \cite{Bhabha11}. 

\begin{figure}[b]
\begin{center}
\resizebox{0.85\columnwidth}{!}{\includegraphics{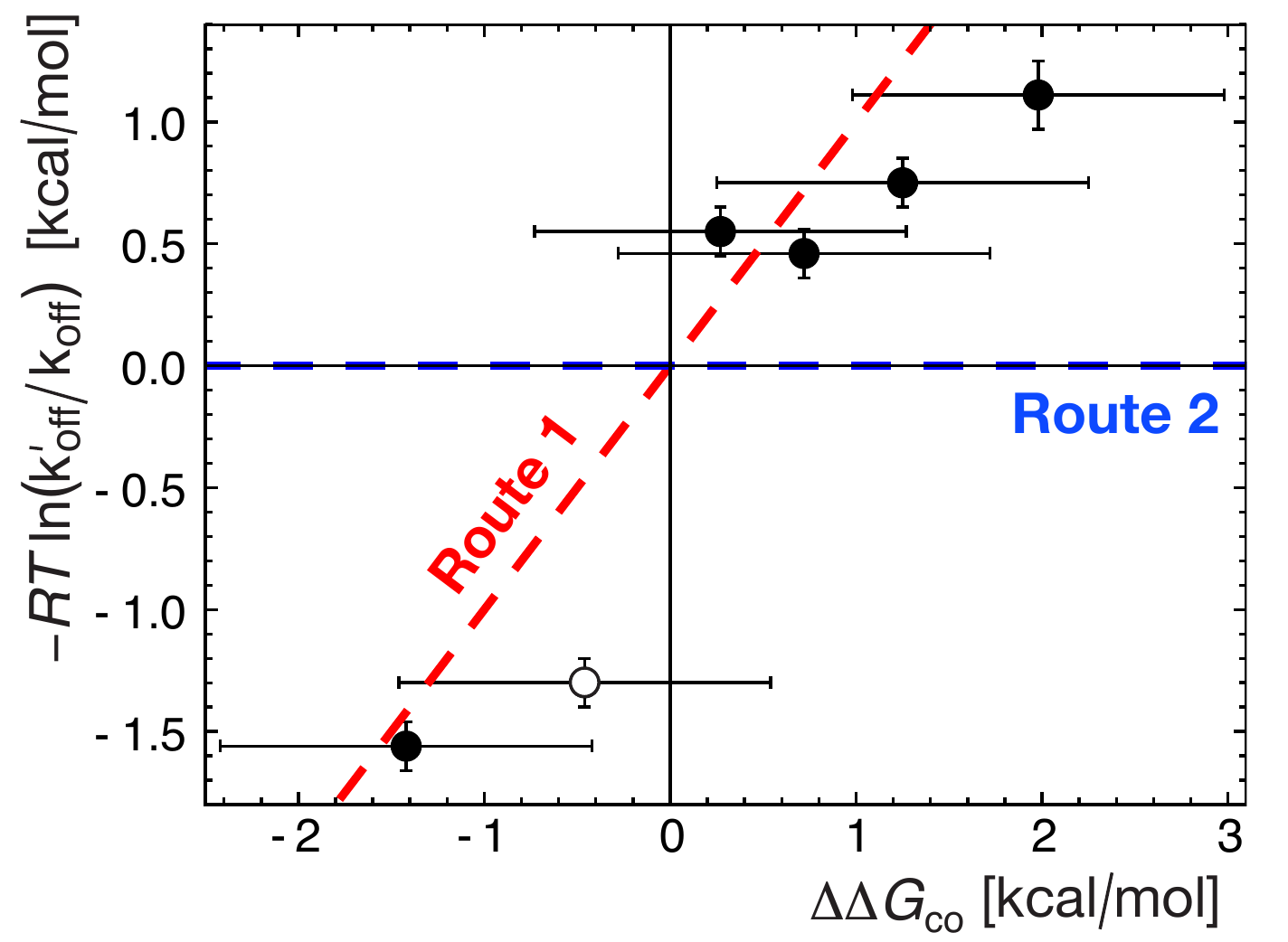}}
\caption{Analysis of experimentally measured product unbinding rates $k_\text{off}^\prime$ and calculated changes $\Delta\Delta G_{co}$ in conformational free-energy differences for six distal mutations of {\em E.~coli} DHFR (see table 2). The data are consistent with the conformational-selection route 1 (red line with slope 1 according to eq.~(\ref{koff1ratio})), but are not consistent with the induced-change route 2 (blue line, see eq.~(\ref{koff2ratio})). The data are also consistent with route 3 if the intermediate conformation of this route is similar to the closed conformation (see eq.~(\ref{koff3ratio})). The data point for the mutant S148A (open circle) has to be excluded from this analysis since the closed conformation of this mutant seems to be the ground-state conformation prior and after product unbinding. In S148A, the product thus presumably unbinds directly from the closed conformation.}
\label{figure_data}
\end{center}
\end{figure}

\begin{figure}[t]
\begin{center}
\resizebox{0.65\columnwidth}{!}{\includegraphics{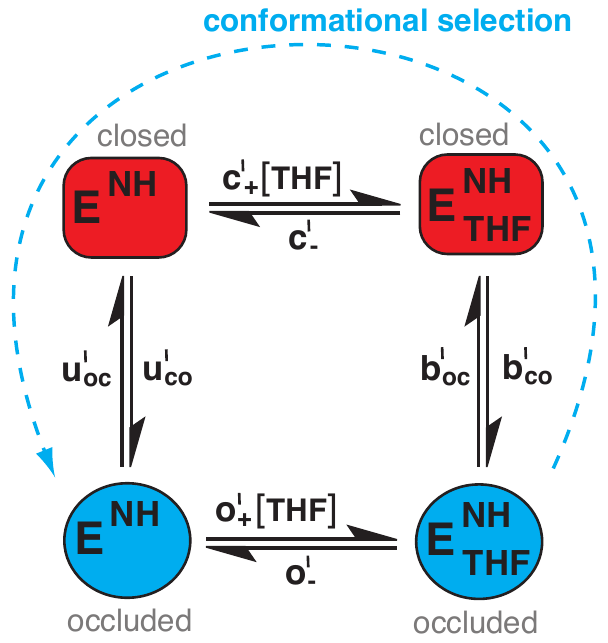}}
\caption{Possible unbinding routes of the product THF from the {\em E.~coli} DHFR mutant G121V, which has the occluded conformation as ground-state conformation prior and after unbinding \cite{Venkitakrishnan04}. The product thus may unbind from this mutant without conformational change, or after a conformational change if the rate $o^\prime_{-}$ for direct unbinding from the occluded conformation is small. The conformational-selection route includes a change from the occluded to the closed conformation prior to product unbinding as on route 1 of the wildtype (see fig.~\ref{figure_routes1and2}), followed by an additional conformational relaxation into the occluded conformation after unbinding. However, in analogy to eq.~(\ref{koff3}) for route 3 of the wildtype, the effective unbinding rate along the conformational-selection route still follows eq.~(\ref{koff1}) for the wildtype route 1 if the relaxation rate $u^\prime_{co}$ into the unbound, occluded ground state is much larger than the rebinding rate $c_{+}^\prime[\text{THF}]$ in the closed conformation, which seems plausible.
}
\label{figure_G121V_product_unbinding}
\end{center}
\end{figure}

The mutation-induced change $\Delta\Delta G_{co}$ of the free-energy difference between the closed and occluded conformation can be calculated as the difference between the stability change of the closed ground-state conformation after product unbinding and the stability change of the occluded ground-state conformation in the product-bound state (see table 2). Within the calculation errors, the $\Delta\Delta G_{co}$ values are proportional to the values of $\ln \left( k_\text{off}^\prime/k_\text{off} \right)$ obtained from the experimentally measured off-rates of the wildtype and mutants (see fig.~\ref{figure_data}), in agreement  with eq.~(\ref{koff1ratio}) for the conformational-selection unbinding route 1, or with eq.~(\ref{koff3ratio}) for route 3 and an intermediate conformation $i$ that is similar to the closed conformation. Our mutational analysis thus indicates that the conformational change from the occluded to the closed conformation occurs, at least to a large extent, prior to product unbinding. The mutation S148A (open circle in fig.~\ref{figure_data}) has to be excluded from this analysis because the product presumably unbinds directly from the closed ground-state conformation of this mutant (see above). The mutation G121V is included in our analysis since the effective unbinding rate along its conformational-selection route (see fig.~\ref{figure_G121V_product_unbinding}) still follows eq.~(\ref{koff1}) for route 1 if the relaxation rate $u^\prime_{co}$ for the additional conformational change into the occluded ground state after unbinding is much larger than the rebinding rate $c_{+}^\prime[\text{THF}]$ in the closed conformation, which seems plausible. 

\begin{table}[h]
\caption{Mutational analysis of the DHFR product unbinding kinetics}
\label{table_product_unbinding}
\begin{center}

\begin{tabular}{l | ccc}
mutation & $k_\text{off}^\prime$ & $-RT\ln (k_\text{off}^\prime/k_\text{off})$  &  $\Delta\Delta G_{co}$ \\
\hline
D122A & $3.5\pm 0.1$ $^{(i)}$  & $0.75\pm 0.10$  & 1.25 \\
D122N & $5.7\pm 0.1$ $^{(i)}$ & $0.46 \pm 0.10$ & 0.72 \\
D122S & $4.9\pm 0.1$ $^{(i)}$ & $0.55 \pm 010$ & 0.27 \\ 
G121V & $1.9\pm 0.3$ $^{(j)}$ & $1.11 \pm 0.14$ & 1.98 \\
M42W & $175$ $^{(k)}$& $-1.56\pm 0.10$ & $-1.42$ \\
S148A & $113\pm 2$ $^{(l)}$ & $-1.30 \pm 0.10$ & $-0.46$ \\
\hline 
\end{tabular}
\end{center}
\end{table}
\vspace*{-0.5cm}
\begin{minipage}[h]{0.95\columnwidth}
\small
$^{(i)}$from ref.~\cite{Miller98c};
$^{(j)}$from ref.~\cite{Cameron97};
$^{(k)}$from ref.~\cite{Rajagopalan02};
$^{(l)}$from ref.~\cite{Miller01}.
-
\small The values for $-RT\ln (k_\text{off}^\prime/k_\text{off})$ have been calculated for the product unbinding rate $k_\text{off}=12.5\pm 2$ s$^{-1}$ of the wildtype \cite{Fierke87} and the given rates $k_\text{off}^\prime$ of the mutants measured at the temperature $T = 25 ^\circ C$ and $\text{pH=7}$. The mutation-induced changes of the free-energy difference between the closed and occluded conformation here is calculated from table 1 as $\Delta\Delta G_{co} = \Delta\Delta G_{cD} - \Delta\Delta G_{oD}$ where  $\Delta\Delta G_{cD} = \Delta\Delta G(\text{E}^\text{NH})$ is the mutation-induced stability change (free energy change relative to the denatured state D) of the closed ground-state conformation after product unbinding,  and $\Delta\Delta G_{oD} = \Delta\Delta G(\text{E}^\text{NH}_\text{THF})$ the stability change of the occluded ground-state conformation with bound product. The units of the $k_\text{off}^\prime$ values are $s^{-1}$, and the units of the values for $-RT\ln (k_\text{off}^\prime/k_\text{off})$ and $\Delta\Delta G_{co}$ are kcal/mol. The estimation of the error of $\Delta \Delta G_{co}$ is complicated by the fact that the mutation-induced free-energy change of the denatured state, which enters the $\Delta\Delta G$ calculation, drops out in the difference since the denatured state is the same for both conformations. We assume that the error for $\Delta \Delta G_{co}$ is similar to the error for $\Delta\Delta G$ and, thus, about 1 kcal/mol (see table 1). This assumption implies similar errors for the mutation-induced free-energy changes of the denatured state and the native conformations in the $\Delta\Delta G$ calculations. 
\end{minipage}

\subsection*{The extended catalytic cycle of wild-type DHFR}

NMR relaxation experiments indicate that wildtype {\em E.~coli} DHFR populates a product-bound excited-state conformation that is rather similar to, but not identical with, the closed ground-state conformation after product unbinding \cite{Boehr06a}. In addition, the experiments indicate that the excited-state conformations in the product-bound state $\text{E}_\text{THF}^\text{NH}$ and in the unbound state  $\text{E}^\text{NH}$ are similar \cite{Boehr06a}, as on route (3) for product unbinding (see eq.~(\ref{route3})). This structural information from the NMR relaxation experiments and our mutational analysis of the previous section point towards a product unbinding route along which most, but not all, of the conformational change from the { occluded to the closed} conformation occurs prior to product unbinding. 

We focus here on the main conformational change from the bound ground state $\text{oE}_\text{THF}^\text{NH}$ to the bound excited state $\tilde{\text{c}}\text{E}_\text{THF}^\text{NH}$ with a conformation similar to the closed conformation, and decompose the product unbinding process into two rather than three steps:
\begin{equation}
\text{oE}_\text{THF}^\text{NH} \;
\begin{matrix} 
\text{\footnotesize $b_{o\tilde{c}}$}\\[-0.05cm]
\text{\Large $\rightleftharpoons$}\\[-0.15cm]
\text{\footnotesize $b_{\tilde{c}o}$} \\[0.05cm]
\end{matrix}
 \; \tilde{\text{c}}\text{E}_\text{THF}^\text{NH} 
\begin{matrix} 
\text{\footnotesize $\tilde{c}_{-}$}\\[-0.05cm]
\text{\Large $\rightleftharpoons$}\\[-0.15cm]
\text{\footnotesize $c_{+}\mbox{[THF]}$} \\[0.05cm]
\end{matrix}
\; \text{cE}^\text{NH}
\label{route1plus}
\end{equation}
The second step from the bound excited state $\tilde{\text{c}}\text{E}_\text{THF}^\text{NH}$ to the unbound state $\text{cE}^\text{NH}$ includes a small conformational change besides the unbinding of the product. We do not further decompose this step since the conformational change during the step appears to be relatively small, and since the conformational transition rates in the unbound state have been determined in NMR relaxation experiments at the rather low temperature of 8$^\circ$C \cite{Boehr06a}. The rates for the the main conformational transition in the product-bound state of the process (\ref{route1plus}) have been determined in NMR relaxation experiments at 27$^\circ$C. These rates are $b_{o\tilde{c}} = 18.5 \pm 1.5$ s$^{-1}$ and $b_{\tilde{c}o} = 735 \pm 45$ s$^{-1}$ \cite{Boehr06a}.

The rates $\tilde{c}_{-}$ and $c_{+}$ for the second step of the unbinding process (\ref{route1plus}) can be determined from the experimentally measured values for the overall rates $k_\text{off} = 12.5 \pm 2$ s$^{-1}$ and $k_\text{on} = 2 \pm 0.2$ $\mu\text{M}^{-1}\text{s}^{-1}$ at 25$^\circ$C \cite{Fierke87}. The unbinding process (\ref{route1plus}) is similar to the conformational-selection unbinding route 1 (see eq.~(\ref{route1})). From eq.~(\ref{koff1}) for the effective off-rate of this route, we obtain
\begin{equation}
\tilde{c}_{-} \simeq \frac{b_{\tilde{c}o}k_\text{off}}{b_{o\tilde{c}} - k_\text{off}} = 1500 \pm 800\text{ s}^{-1}
\label{cmin}
\end{equation}
with the experimental values for $b_{o\tilde{c}}$,  $b_{\tilde{c}o}$, and $k_\text{off}$ given above. From eq.~({\ref{kon1}}), we obtain
\begin{equation}
c_{+} \simeq \frac{(b_{\tilde{c}o} + \tilde{c}_{-}) k_\text{on}}{b_{\tilde{c}o}} = 6 \pm 2 \;\;\mu\text{M}^{-1}\text{s}^{-1}
\label{cplus}
\end{equation}
with the experimental values for $b_{c\tilde{o}}$ and $k_\text{on}$ and the value for $\tilde{c}_{-}$ from eq.~(\ref{cmin}). The resulting value of $c_{+}$ justifies the assumption $b_{\tilde{c}o} \gg c_{+}[\text{THF}]$ made in the derivation of eq.~({\ref{kon1}}) since typical product concentrations [THF] are around 10 $\mu$M in {\em E.\ coli} \cite{Fierke87b}. 

Our decomposition of the product unbinding step into the two substeps of eq.~(\ref{route1plus}) is included in the extended catalytic cycle shown in fig.~\ref{figure_WT_cycle_extended}. Together with the decomposition of the catalytic step in eq.~(\ref{catalysis_DHFR}), the extended cycle specifies the ordering of events in the coupling of the main conformational transition between the closed and occluded active-site loop conformations with binding and catalysis.

\section*{DISCUSSION}

%
\subsection*{Role of the conformational changes in wildtype  {\em E.~coli} DHFR}

In contrast to {\em E.~coli} DHFR,  human and other vertebrate DHFRs only populate the closed conformation of the active-site loop along their catalytic cycle \cite{Davies90,Bhabha11}. This raises the question why there are conformational changes between a closed and an occluded active-site loop along the catalytic cycle of {\em E.~coli} DHFR. If the conformational changes were `unfavorable', they could be `eliminated' by mutations such as S148A that strongly destabilize the occluded conformation relative to the closed conformation. Similar to human DHFR, the mutant S148A of {\em E.~coli} DHFR only populates the closed conformation \cite{Bhabha11}, which is the required conformation for the chemical step.

In contrast to human and other vertebrate cells, the internal pH of  {\em E.~coli} bacteria can vary from values below 5  \cite{Richard04} to peak values possibly larger than 8.7 \cite{Kroll83}, depending on the external pH of the surrounding medium. At low pH values, the overall catalytic rate of the extended catalytic cycle shown in fig.~\ref{figure_WT_cycle_extended} is clearly limited by the conformational change prior to product unbinding. The conformational change with excitation rate 18.5  s$^{-1}$ reduces the effective product unbinding rate to 12.5 s$^{-1}$ (see eq.~(\ref{koff1})), which is almost two orders of magnitude smaller than the maximal forward rate 950  s$^{-1}$  of the chemical step attained at low pH values. At pH values close to the preferred internal pH 7.8 of  {\em E.~coli} \cite{Richard04}, in contrast, the forward rate $k_f$ of the chemical step is of the same order of magnitude as the product unbinding rate (see eq.~(\ref{kf(pH)})). The change from the occluded to the closed conformation prior to product unbinding thus balances the strong pH dependence of the chemical reaction by limiting the overall catalytic rate at low pH to values of the same order of magnitude as the catalytic rate at the preferred pH of  {\em E.~coli}.

Another difference between {\em E.~coli} and human and other vertebrate cells concerns the relative concentrations of NADPH (NH) and NADP$^+$ (N$^+$). In prokaryotic cells such as {\em E.~coli}, the concentrations of NH and N$^+$ are comparable \cite{Miller01}. In eukaryotic cells, in contrast, the concentration of N$^+$ is typically smaller than 1\% of the NH concentration. A consequence of the comparable concentrations of NH and N$^+$ is that {\em E.~coli} DHFR works relatively close to the chemical equilibrium at pH values  larger than the preferred pH value 7.8. In equilibrium, the product of all the rates in the forward direction of the catalytic cycle   
is equal to the product of all rates in the backward direction \cite{Hill89}. This equilibrium condition leads to 
\begin{equation}
1.3\cdot 10^{11-\text{pH}}\,[\text{DHF}][\text{NH}]= [\text{THF}][\text{N}^{+}]
\label{equilibrium}
\end{equation}
for the forward rate $k_f$ of the catalytic step given in eq.~(\ref{kf(pH)}) and the reverse rate  $k_r =0.55$ s$^{-1}$ at high pH. For the exemplary concentrations [DHF] = 0.3  $\mu$M, [THF] = 13 $\mu$M, [NH] = 1  mM, and [N$^{+}$] = 1.5 mM  of Benkovic and coworkers \cite{Fierke87b}, equilibrium is attained at pH = 9.3 according to eq.~(\ref{equilibrium}). The pH value at which equilibrium is attained is even lower for smaller substrate concentrations or higher product concentration than in this example. Fluctuations of the substrate and product concentrations thus may lead to a reversal of the catalytic cycle of DHFR at pH values larger than the preferred pH of  {\em E.~coli}.  

In reverse direction, the catalytic rate of {\em E.~coli} DHFR is limited by the low rate $k_r$ of the reverse catalytic step. On the extended catalytic cycle shown in  fig.~\ref{figure_WT_cycle_extended}, the low value of $k_r$ is a consequence of  the conformational change that precedes the chemical step in reverse direction. According to eq.~(\ref{kmin}), the actual rate of the reverse chemical step in the closed conformation is $k_{-}\simeq 38$ s$^{-1}$ at high pH, and the overall low reverse rate $k_r \simeq 0.55$ s$^{-1}$ results from the change between the occluded ground-state conformation and closed excited-state conformation with bound products THF and N$^+$ prior to the reverse chemical step. This conformational change thus limits the negative effects of a possible reversal of the catalytic cycle by strongly reducing the reverse catalytic rates.

The conformational changes of {\em E.~coli} DHFR thus may provide robustness of the catalytic process against pH variations and changes in the substrate, product and cofactor concentrations. In the forward direction of the catalytic cycle, the conformational change prior to product unbinding prevents strong deviations of the overall catalytic rate for pH values below the preferred pH of  {\em E.~coli}. And at high pH, the conformational change during the catalytic step prevents high catalytic flux in the reverse direction.

\subsection*{Stabilization of the occluded conformation can lead to low rates of the catalytic step}

Active-site remote mutations  such as G121V in the FG loop  of {\em E.~coli} DHFR (see fig.~\ref{figure_structures}) indicate that even distal residues in an enzyme can play important catalytic roles.  The forward rate for the catalytic step of the mutant G121V is $k_f^\prime \simeq 1.4\text{~s}^{-1}$ at pH 7 and, thus, significantly smaller than the wildtype rate $k_f \simeq 220\text{~s}^{-1}$ at the same pH. We argue here that this strong decrease in the forward rate for the catalytic step is related to a stabilization of the occluded conformation relative to the closed conformation. NMR experiments with the mutant G121V indicate that the occluded conformation is the ground-state conformation in all five intermediates along the catalytic cycle \cite{Venkitakrishnan04}. Since hydride transfer from NH to DHF is only possible in the closed conformation, the catalytic step of the mutant G121V requires an excitation into the closed conformation prior to the chemical reaction, and a relaxation into the occluded conformation after the reaction:
\begin{equation}
\text{oE}_\text{DHF}^\text{NH} \;
\begin{matrix} 
\text{\footnotesize $s_{oc}^\prime$}\\[-0.05cm]
\text{\Large $\rightleftharpoons$}\\[-0.15cm]
\text{\footnotesize $s_{co}^\prime$} \\[0.05cm]
\end{matrix}
\; \text{cE}_\text{DHF}^\text{NH} \;
\begin{matrix} 
\text{\footnotesize $k_{+}^\prime$}\\[-0.05cm]
\text{\Large $\rightleftharpoons$}\\[-0.15cm]
\text{\footnotesize $k_{-}^\prime$} \\[0.05cm]
\end{matrix}
 \; \text{cE}_\text{THF}^\text{N+} \;
 \begin{matrix} 
\text{\footnotesize $k_{co}^\prime$}\\[-0.05cm]
\text{\Large $\rightleftharpoons$}\\[-0.15cm]
\text{\footnotesize $k_{oc}^\prime$} \\[0.05cm]
\end{matrix}
\; \text{oE}_\text{THF}^\text{N+}
\label{catalytic_step_G121V}
\end{equation}
In contrast, the catalytic step of the wildtype only involves a conformational relaxation after the chemical substep (see eq.~(\ref{catalysis_DHFR})). The effective forward rate for this catalytic 3-step process of the mutant G121V is (see eq.~(\ref{kAD}))
\begin{eqnarray}
k_f^\prime &\simeq& \frac{s^\prime_{oc} k^\prime_{+} k^\prime_{co}}{s^\prime_{co} (k^\prime_{-} + k^\prime_{co})+ k^\prime_{+} k^\prime_{co} } \nonumber\\
& \simeq& \frac{s_{oc}^\prime k_{+}^\prime}{s_{co}^\prime + k_{+}^\prime} \text{~~~for~} k^\prime_{co} \gg k^\prime_{-} \nonumber\\
&\simeq& \frac{s_{oc}^\prime}{s_{co}^\prime} k_{+}^\prime  = K_s^\prime k_{+}^\prime  \text{~~~for~} s_{co}^\prime \gg k_{+}^\prime
\label{kf_G121V}
\end{eqnarray}
with the constant $K_s^\prime = s_{oc}^\prime/s_{co}^\prime$ for the conformational equilibrium of the substrate-bound states. 
We have assumed here that the relaxation rate $k_{co}^\prime$ into the product-bound ground state $\text{oE}_\text{THF}^\text{N+}$ is much larger than $k_{-}^\prime$, as for the wildtype (see fig.~\ref{figure_WT_cycle_extended}). In addition, we have assumed that the relaxation rate $s_{co}^\prime$ into the substrate-bound ground state is much larger than $k_{+}^\prime$, which is reasonable at least for intermediate pH values (see eqs.~(\ref{kf(pH)}) and (\ref{kf})).

The small forward rate $k_f^\prime$  for the catalytic step of the mutant G121V is a consequence of the fact that the conformational equilibrium constant $K_s^\prime $ in eq.~(\ref{kf_G121V}) is much smaller than 1 since the occluded conformation is the ground-state conformation. According to eq.~(\ref{kf_G121V}), the forward rate $k_f^\prime$ of the mutant G121V is the product of the equilibrium constant  $K_s^\prime$ and the rate  $k_{+}^\prime$ for the chemical substep. In contrast, the forward rate $k_f$ for the catalytic step of the wildtype is identical with the rate $k_f$ of the chemical substep (see eq.~(\ref{kf})). We would like to emphasize that eq.~(\ref{kf_G121V}) is in agreement with the observed kinetic isotope effect for the forward rate of G121V \cite{Cameron97,Wang06a,Wang06b,Swanwick06} since the rate $k_f^\prime$ is proportional to the rate $k_{+}^\prime$ of the actual chemical step. The catalytic scheme (\ref{catalytic_step_G121V}) is also in agreement with pre-steady-state kinetic experiments, which indicate a conformational change prior to the chemical reaction with an excitation rate of $s_{oc}^\prime \simeq 3.5 \text{~s}^{-1}$ \cite{Cameron97}. Similar to G121V, the strong reduction in the forward rate of the catalytic step observed for other distal mutations such as D122A, D122N, and D122S \cite{Miller98c} may also results from a stabilization of the occluded conformation.

We have attributed here strong reductions in the forward rate of the catalytic step caused by distal mutations to changes in the conformational equilibrium. For mutations within the active-site loop (Met20 loop), a different, dynamic mechanism has been recently proposed by Bhabha et al.~\cite{Bhabha11}. The mutant N23PP and the double mutant  N23PP/S148A only populate the closed state of the active-site loop, but reduce the forward rate $k_f$ for the catalytic step by a factor of 15. This reduction is traced back to a stiffening of the active-site loop by the mutation N23PP, which dampens loop fluctuations that seem to be required for catalysis in the closed conformation\cite{Bhabha11}. 

\section*{CONCLUSIONS}

In this article, we have presented extended catalytic cycles for the enzyme DHFR from {\em E.~coli} that specify the ordering of chemical and physical steps, including conformational changes. The extended cycles were derived in a general theoretical framework that provides a link between the conformational transition rates from dynamic NMR experiments and the effective binding and catalytic rates measured in relaxation experiments. Our theoretical framework allows to distinguish between {\em induced} conformational changes that occur predominantly {\em after} a binding/unbinding or chemical process, and {\em selected} conformational changes that occur predominantly {\em prior to} such processes. In addition, the framework helps to explain the effect of mutations on the rates of binding and catalysis. Our approach is rather general and applicable to other proteins and  experimental methods that probe the conformational or binding dynamics.

\section*{APPENDIX}

\subsection*{Relaxation rates of the three-state process}

The catalytic process of eq.~(\ref{catalysis_DHFR}) and the routes 1 and 2  for product unbinding of {\em E.~coli} DHFR follow the general 3-state reaction scheme 
\begin{equation}
A \;
\begin{matrix} 
\text{\footnotesize $k_{AB}$}\\[-0.05cm]
\text{\Large $\rightleftharpoons$}\\[-0.15cm]
\text{\footnotesize $k_{BA}$} \\[0.05cm]
\end{matrix}
 \; B \;
\begin{matrix} 
\text{\footnotesize $k_{BC}$}\\[-0.05cm]
\text{\Large $\rightleftharpoons$}\\[-0.15cm]
\text{\footnotesize $k_{CB}$} \\[0.05cm]
\end{matrix}
\; C 
\label{3state}
\end{equation}
The time-dependent probabilities $P_A(t)$, $P_B(t)$, and $P_C(t)$ of the three states A, B, and C in this scheme are governed by the master equations
\begin{eqnarray*}
\frac{d P_A(t)}{dt} &=& k_{BA} P_B(t) - k_{AB} P_A(t)  \label{3_one}\\
\frac{d P_B(t)}{dt} &=& k_{AB} P_A(t) - (k_{BA} + k_{BC})P_B(t) + k_{CB} P_C(t) \\
\frac{d P_C(t)}{dt} &=& k_{BC} P_B(t) - k_{CB} P_C(t) \label{3_three}
\end{eqnarray*}
where $k_{ij}$ denotes the rate from state $i$ to state $j$. The three equations can be written in the matrix form
\begin{equation}
\frac{d \boldsymbol{P}(t)}{dt} = - \boldsymbol{W} \boldsymbol{P}(t)  \label{matrix_equation}
\end{equation}
with $\boldsymbol{P}(t)=\left(P_A(t),P_B(t),P_C(t)\right)$ and 
\begin{equation}
\boldsymbol{W} = 
\begin{pmatrix} 
k_{AB} & - k_{BA} & 0\\
- k_{AB} & k_{BA} + k_{BC} & -k_{CB}\\
0 & -k_{BC} & k_{CB}
\end{pmatrix}
\end{equation}
The general solution of eq.~(\ref{matrix_equation}) has the form \cite{Kampen92}
\begin{equation}
\boldsymbol{P}(t) = c_o \boldsymbol{Y}_o + c_1 \boldsymbol{Y}_1 e^{-k_1 t} + c_2 \boldsymbol{Y}_1 e^{-k_2 t}
\label{gensol}
\end{equation}
where $\boldsymbol{Y}_o$,  $\boldsymbol{Y}_1$, and $\boldsymbol{Y}_2$ are the three eigenvectors of the matrix $\boldsymbol{W}$, and $k_1$ and $k_2$ are the two positive eigenvalues 
\begin{widetext}
\begin{eqnarray}
k_{1} = \frac{1}{2} \Big(k_{AB} + k_{BA} + k_{BC} + k_{CB} 
- \sqrt{(k_{AB} + k_{BA} + k_{BC} + k_{CB})^2 - 4(k_{AB}(k_{BC}+k_{CB})+k_{CB}k_{BA})}\Big) 
\label{k1}\\
k_{2} = \frac{1}{2} \Big(k_{AB} + k_{BA} + k_{BC} + k_{CB} 
+ \sqrt{(k_{AB} + k_{BA} + k_{BC} + k_{CB})^2 - 4(k_{AB}(k_{BC}+k_{CB})+k_{CB}k_{BA})}\Big) 
\label{k2}
\end{eqnarray}
\end{widetext}
of $\boldsymbol{W}$ with corresponding eigenvectors $\boldsymbol{Y}_1$ and $\boldsymbol{Y}_2$.  The smaller eigenvalue $k_1$ is the {\em dominant relaxation rate} on long time scales. This eigenvalue corresponds to the relaxation rate
\begin{equation}
k_\text{obs} \simeq k_1
\label{kobs}
\end{equation}
observed in experiments.

\subsection*{Effective rates of the three-state process}  

A general {\em effective rate} $k_{A\to C}$ from state A to state C can be defined by considering an adsorbing state C without probability backflow into state B. An absorbing state C with $k_{CB} = 0$ corresponds, e.g., to a situation in which the probability flow from state C into another state on a chemical cycle is much faster than the backflow into B. The effective rate from A to C is the dominant relaxation rate $k_1$ for $k_{CB} = 0$:
\begin{equation}
k_{A\to C} = k_1\big |_{k_{CB} = 0}
\end{equation}
The effective rate $k_{A\to C}$  depends on the three rates $k_{AB}$, $k_{BA}$, and $k_{BC}$ (see eq.~(\ref{k1})).
In limiting cases in which one of these rates is much smaller than another rate, we obtain
\begin{equation}
k_{A\to C} \simeq
\begin{cases}
\displaystyle \frac{k_{AB}k_{BC}}{k_{BA}+k_{BC}} \text{~~~for~} k_{AB} \ll k_{BA} \text{~or~} k_{AB} \ll k_{BC}  \\[0.3cm]
\displaystyle\frac{k_{AB}k_{BC}}{k_{AB}+k_{BA}} \text{~~for~} k_{BC} \ll k_{BA} \text{~or~} k_{BC} \ll k_{AB} \\[0.3cm]
\displaystyle\min(k_{AB}, k_{BC}) \text{~for~} k_{BA}\ll k_{AB} \text{~or~} k_{BA} \ll k_{BC}
\end{cases}
\label{kAC}
\end{equation}

An important special case of the process (\ref{3state}) is the case in which the equilibrium probability of state B is significantly smaller than the equilibrium probabilities of A and C. In this case, we have $k_{AB}\ll k_{BA}$ and $k_{CB}\ll k_{BC}$, and the relaxation rate (\ref{kobs})  is
\begin{equation}
k_\text{obs} \simeq \frac{k_{AB} k_{BC} + k_{BA} k_{CB}}{k_{BA} + k_{BC}} = k_{A\to C} + k_{C\to A}
\label{rate3state}
\end{equation}
with the effective rates
\begin{equation}
k_{A \to C} \simeq \frac{k_{AB} k_{BC} }{k_{BA} + k_{BC}}  \text{~~for~} k_{AB}\ll k_{BA}
\end{equation}
and 
\begin{equation}
k_{C \to A} \simeq \frac{k_{BA} k_{CB}}{k_{BA} + k_{BC}} \text{~~for~} k_{CB}\ll k_{BC}
\end{equation}
obtained from eq.~(\ref{kAC}). In the special case of a low-probability intermediate state B,  the dominant relaxation rate (\ref{rate3state}) is also obtained in the steady-state approximation for the 3-state process (\ref{3state}). The steady-state approximation is based on the assumption $d P_B(t)/dt= 0$, i.e.~on a negligible change of the probability for the state B in the set of equations (\ref{3_one}) to (\ref{3_three}).  In general, however, the dominant relaxation rate $k_\text{obs}$ is not equal to $k_{A \to C}+k_{C \to A}$.

{ Our derivation of the eqs.~(\ref{kmin}), (\ref{kf}), (\ref{koff1}), and (\ref{kon1}) for the effective rates of the catalytic step and the product unbinding step along the catalytic cycle of {\em E.~coli} is based on eq.~(\ref{kAC}) and, thus, on relaxation rates. This derivation is in line with the enzyme kinetics measurements, in which the rates for these steps are determined from pre-steady-state or relaxation experiments \cite{Fierke87}. Alternatively, the effective rates for these steps can be derived from results for the steady-state rate of the enzymatic cycle (T. R. Weikl, unpublished).
 }

\subsection*{Effective rates of a four-state process}

Route 3 for product unbinding of {\em E.~coli} DHFR follows the general four-state reaction scheme 
\begin{equation}
A \;
\begin{matrix} 
\text{\footnotesize $k_{AB}$}\\[-0.05cm]
\text{\Large $\rightleftharpoons$}\\[-0.15cm]
\text{\footnotesize $k_{BA}$} \\[0.05cm]
\end{matrix}
\; B \;
\begin{matrix} 
\text{\footnotesize $k_{BC}$}\\[-0.05cm]
\text{\Large $\rightleftharpoons$}\\[-0.15cm]
\text{\footnotesize $k_{CB}$} \\[0.05cm]
\end{matrix}
 \; C \;
 \begin{matrix} 
\text{\footnotesize $k_{CD}$}\\[-0.05cm]
\text{\Large $\rightleftharpoons$}\\[-0.15cm]
\text{\footnotesize $k_{DC}$} \\[0.05cm]
\end{matrix}
\; D 
\label{4state}
\end{equation}
The probability evolution of the four states is governed by the master equation set
\begin{eqnarray*}
\frac{d P_A(t)}{dt} &=& k_{BA} P_B(t) - k_{AB} P_A(t)  \label{4_one}\\
\frac{d P_B(t)}{dt} &=& k_{AB} P_A(t) - (k_{BA} + k_{BC})P_B(t) +k_{CB}P_C(t)\\
\frac{d P_C(t)}{dt} &=& k_{BC} P_B(t)  - (k_{CB} + k_{CD})P_C(t) + k_{DC}P_D(t)\\
\frac{d P_D(t)}{dt} &=& k_{CD} P_C(t) - k_{DC}P_D(t)  \label{4_four}
\end{eqnarray*}
These equations can be written in the form of the matrix equation (\ref{matrix_equation}) with $\boldsymbol{P}(t)=\left(P_A(t),P_B(t),P_C(t),P_D(t)\right)$ and 
\begin{equation}
\boldsymbol{W} = 
\begin{pmatrix} 
k_{AB} & - k_{BA} & 0 & 0\\
- k_{AB} & k_{BA} + k_{BC} & -k_{CB} & 0\\
0 & -k_{BC} & k_{CB} + k_{CD} & -k_{DC} \\
0 & 0 & -k_{CD} & k_{DC} 
\end{pmatrix}
\label{4matrix}
\end{equation}
The three relaxation rates of the four-state process are the three positive eigenvalues of the matrix $\boldsymbol{W}$. 

The states B and C of the process (\ref{4state}) correspond to the excited states $iE_\text{THF}^\text{NH}$ and $iE^\text{NH}$ along route 3, which exchange with the ground states with rates $b_{oi}\ll b_{io}$ and $u_{ci}\ll u_{ic}$.  We focus here on the corresponding case $k_{AB}\ll k_{BA}$ and $k_{DC}\ll k_{CD}$. In this case, the intermediate state B has a low equilibrium probability compared to A, and the state C a low probability compared to D. This makes the steady-state approximation plausible, which assumes $dP_B(t)/dt = 0$ and $dP_C(t)/dt = 0$ for the intermediate states. Within this approximation, the four equations (\ref{4_one}) to (\ref{4_four}) can be solved easily, which leads to the relaxation rate
\begin{equation} 
 k_\text{obs} \simeq \frac{k_{AB} k_{BC} k_{CD}+k_{BA} k_{CD} k_{DC}}{k_{BA} k_{CB} + k_{BA} k_{CD}+ k_{BC} k_{CD} }
\label{kfour}
\end{equation} 
For $k_{AB}\ll k_{BA}$ and $k_{DC}\ll k_{CD}$, the approximate expression (\ref{kfour}) is an excellent agreement with numerical results for the dominant relaxation rate of the process (\ref{4state}), which is the smallest nonzero eigenvalue of the matrix (\ref{4matrix}).

Similar to the three-state process with low-probability intermediate B, the relaxation rate $k_\text{obs}$ is the sum of the effective rate 
\begin{equation}
k_{A \to D}\simeq \frac{k_{AB} k_{BC} k_{CD}}{k_{BA} k_{CB} + k_{BA} k_{CD}+ k_{BC} k_{CD} }
\label{kAD}
\end{equation}
from A to D and the effective backward rate
\begin{equation}
k_{D \to A} \simeq \frac{k_{BA} k_{CD} k_{DC}}{k_{BA} k_{CB} + k_{BA} k_{CD}+ k_{BC} k_{CD} }
\end{equation}
from D to A. The effective rate form A to D is the dominant relaxation rate for $k_{CD}=0$, i.e.~for an absorbing state D without probability backflow into C. The effective rate from D to A is the dominant relaxation rate for $k_{AB}=0$, i.e.~for an absorbing state A.

\end{document}